\documentclass[a4paper,12pt]{article}
\usepackage{amsmath}
\usepackage{amsfonts}
\usepackage{amssymb}
\usepackage{latexsym}
\usepackage{epsfig}
\usepackage{graphicx}
\usepackage{oldgerm}
\usepackage{theorem}
\usepackage{bm}

\usepackage[hang,small,bf]{caption}
\usepackage[subrefformat=parens]{subcaption}
\def\N{\mathbb{N}}

\def\bra{\langle}
\def\ket{\rangle}

\theorembodyfont{\itshape}

\newcommand{\SSC}[1]{\section{#1}\setcounter{equation}{0}}

\newfont{\bg}{cmr10 scaled\magstep4}

\newcommand{\bigzerou}{\smash{\lower1.7ex\hbox{\bg 0}}}

\begin{document}

\title{
\bf 
Switching particle systems for foraging ants \\
showing phase transitions in path selections
}
\author{
Ayana Ezoe$^1$, 
Saori Morimoto$^1$,
Yuya Tanaka$^1$, 
Makoto Katori$^1$
\footnote{$^1$
Department of Physics,
Faculty of Science and Engineering,
Chuo University, 
Kasuga, Bunkyo-ku, Tokyo 112-8551, Japan;
makoto.katori.mathphys@gmail.com
} 
\\ and
Hiraku Nishimori$^2$
\footnote{$^2$
Meiji Institute for Advanced Study of Mathematical Sciences, 
Meiji University, 
Nakano, Nakano-ku, Tokyo 164-8525, Japan;
e-mail: nishimor2@meiji.ac.jp
}
}

\date{8 April 2024}
\pagestyle{plain}
\maketitle

\begin{abstract}
Switching interacting particle systems 
studied in probability theory are
the stochastic processes of 
hopping particles on a lattice
made up of slow 
and fast particles, where
the switching between these types of
particles occurs randomly at a given transition rate.
This paper explores how
such stochastic processes involving multiple particles 
can model group behaviors of ants.
Recent experimental research by the last author's group
has investigated how ants switch 
between two types of 
primarily relied cues to select foraging paths 
based on the current situation. 
Here, we propose a discrete-time interacting random walk
model on a square lattice, incorporating
two types of hopping rules. 
Numerical simulation results demonstrate
global changes in selected homing paths, 
transitioning 
from trailing paths of the `pheromone road' 
to nearly optimal paths 
depending on the switching parameters.
By introducing two types of order parameters
characterizing the dependence
of homing duration distributions on switching parameters, 
we discuss these global changes as phase transitions
in ant path selections.
We also study 
critical phenomena associated with 
continuous phase transitions.

\vskip 0.2cm



\end{abstract}
\vspace{3mm}

\clearpage

\SSC
{Introduction}
\label{sec:introduction}

For reaction-diffusion systems
with \textit{double diffusivity}, 
the interacting particle systems 
in one dimension were introduced 
as microscopic models, 
in which we have two types of particles; 
\textit{fast particles} and \textit{slow particles} \cite{Flo22}.
Fast particles hop at rate 1 and
slow particles at rate $\epsilon \in [0, 1)$. 
Switching between two types of particles
occurs at rate $\gamma \in (0, \infty)$. 
An alternative view of the models is to consider
two layers of particles such that
the hopping rate for the bottom layer is 1,
whereas that for the top layer is $\epsilon <1$,
and to let particles change layers at rate $\gamma$. 
It is noteworthy that
the motivation for studying such 
switching particle systems
came from population genetics \cite{Flo22}. 
Individuals living in colonies can behave
either \textit{active} or \textit{dormant}. 
The active individuals undergo clonal reproduction
by resampling within a colony
and migrate between colonies via hopping.
They can become dormant and then
resampling and migration are suspended until
they become active again. 
Dormant individuals reside in \textit{seed banks}.
A useful review of the important roles of dormancy 
in life sciences was provided by \cite{Len12}.
In switching interacting particle systems,  
the active (resp.~dormant) individuals
are represented by fast (resp.~slow) particles.
Other mathematical models and their analyses
for population dynamics with seed banks, see
\cite{dH22a,dH22b,Gre22,Gre23}.

This study aims to show the usefulness of the notion of 
switching interacting particle systems
for modeling group behaviors
for ant foraging \cite{HW90,Nis15}. 
The group behaviors of so-called \textit{eusocial insects}
(\textit{e.g.} ants, bees, and termites) have been
of interest not only in biology \cite{HW90}
but also in statistical physics as 
many-body phenomena in systems consisting of
interacting, permanently moving units. 
Although persistent motion is a common feature of
living systems, recently several physical and chemical 
systems comprising interacting
\textit{self-propelled} units have been reported.
\textit{Interacting self-propelled particle systems}
exhibit richer phenomena in their collective motion
than the thermal or chemical fluctuation phenomena
in equilibrium systems. 
Study of collective motions of self-propelled particles
in both living and non-living worlds
is now one of the most active and fruitful topics
in non-equilibrium statistical physics \cite{VZ12}.
Through direct and indirect communication, ants
in colonies share various types of information
and exhibit highly organized group behaviors
which enable them to perform complex tasks \cite{HW90}.
Foraging is one of the most intensively studied subjects
in such interesting group behaviors \cite{Nis15,VZ12}.

\subsection{Switching particle systems
modeling individual switching behaviors}
\label{sec:switching}
\textit{Recruit pheromone} is a chemical substance 
or mixture released by an organism that influences
the behaviors of other individuals.
Various species of ants establish lengthy foraging
trails through a positive feedback process
involving the secretion and tracking of 
\text{recruit pheromone},
enabling them to efficiently shuttle
between the nest and food sources.
However, a recent combined study of experiments
and mathematical modeling for a species of
garden ant suggested 
a situation-dependent switch of the primarily
relied cues from chemical ones 
(\textit{e.g.} recruit pheromone) 
to other ones, most likely visual cues. 
Here, visual cues refer to landmarks, 
the polarization angle of the sun or moon,
textures of the edges of crowded plants or woods, 
and other visual stimuli \cite{Nis15}.

To elucidate the sudden changes of
path patterns of ant trails 
which were experimentally observed 
depending on geometric configuration
of the nest and food source on a plane
(see Section~\ref{sec:experiment} below), 
Ogihara \textit{et al.} proposed a multi-agent
model \cite{Nis15}. 
In this model, foraging ants are depicted as 
biased random walkers on a lattice
with their probabilities of hopping to 
nearest neighboring sites 
weighted based on the evolving pheromone field
over time.
The local intensity of the pheromone field 
increases through pheromone secretion by exploring ants
and homing ants (with the pheromone field
diminishing due to constant evaporation).
In addition to such a 
\textit{pheromone-mediated walk} of ants,
a \textit{visual-cues-mediated walk} is considered
for homing ants, 
which aims to achieve an optimal path to the nest. 
They introduced a parameter $\alpha$ for each ant at each time, 
which measures the conflict between 
the optimal path and the redundant path
made by pheromone-mediated walk \cite{Nis15}.
To make switching from
the pheromone-mediated walk
to the visual-cues-mediated walk,
they assumed a critical value 
$\alpha_{\rm c}$, and
when $\alpha > \alpha_{\rm c}$ the
switching occurs and the direction of
the hopping is aligned toward the nest. 

In this paper, we propose another mathematical model
for the selection of foraging paths by ants
inspired by the concept of switching interacting 
particle systems \cite{Flo22}.
We consider 
a discrete-time interacting random walk model
of hopping particles on a square lattice
(a cellular automaton model) \cite{CD98,MD99}.
The ants exploring outward from the nest to the food source
are depicted as random walkers
during the \textit{exploring period},
while those returning from the food source to nest
are portrayed as random walkers during the
\textit{homing period}. 
We regard the exploring and homing ants
following the pheromone-mediated walk
as slow particles, and 
the homing ants utilizing the 
visual-cues-mediated walk as fast particles:
\begin{align}
\mbox{pheromone-mediated walk} 
&\iff \mbox{slow particle}, 
\nonumber\\
\mbox{visual-cues-mediated walk}
&\iff \mbox{fast particle}.
\label{eqn:correspond}
\end{align}
During the exploring period,
all particles are slow, whereas
in the homing period,
transitions can occur
from slow particles to fast particles.
These fast particles, representing 
``pioneering ants'', establish shorter paths
to the nest. 
However, pioneering ants may face a
\textit{higher risk encountering enemies} compared to the ants
following previously established paths. 
Therefore, fast particles should revert to
slow particles to mitigate this risk. 
The switching between 
the two types of particle behavior occurs 
randomly at each time step as
\begin{align}
&\mbox{slow particle $\Longrightarrow$ fast particle 
\quad with probability $\gamma_{\rm sf}$}, 
\nonumber\\
&\mbox{fast particle $\Longrightarrow$ slow particle 
\quad with probability $\gamma_{\rm fs}$}.
\label{eqn:gamma}
\end{align}
We will investigate how the global changes in homing paths
depend on the parameters $\gamma_{\rm sf}$ 
and $\gamma_{\rm fs}$.
Notice that, since here we consider a discrete-time stochastic
model, $\gamma_{\rm sf}$ and $\gamma_{\rm fs}$ are
not transition rates but transition probabilities taking
values in $[0, 1]$. 

\subsection
{Experiments suggesting phase transitions}
\label{sec:experiment}

Now, we provide a concise overview of the 
experiments on forage path selection by ants
conducted by the research group led by the last author of
this paper (the Nishimori group).
These experiments provided 
clear evidence of the switching phenomenon
in the primarily relied cues used by ants 
from recruit pheromone to visual information
\cite{Nis15}.

They utilized \textit{Lasius Japonicus},
a species of garden ant found in Japan,
raised in plastic boxes measuring
$23 \times 12 \times 3.5$ cm equipped with 
special floors and walls covered in plaster
to maintain internal humidity.
Each box accommodated approximately 200 to 300 ants
and was shielded by a black plate to shield the ants from light.
Recruit pheromone was extracted
from the remaining ants collected from the
same natural colony using column chromatography. 
Experimental processes were recorded by a video camera. 
For further details on the setup, refer to the original paper \cite{Nis15}.

In the box, a food source was placed at a separate 
position from the nest.
They created a conflicting scenario where the
homing directions for ants from the food source, 
as indicated by chemical cues and visual cues, diverged
from each other as follows:

\begin{figure}[ht]
\begin{center}
\includegraphics[width=0.8\textwidth]{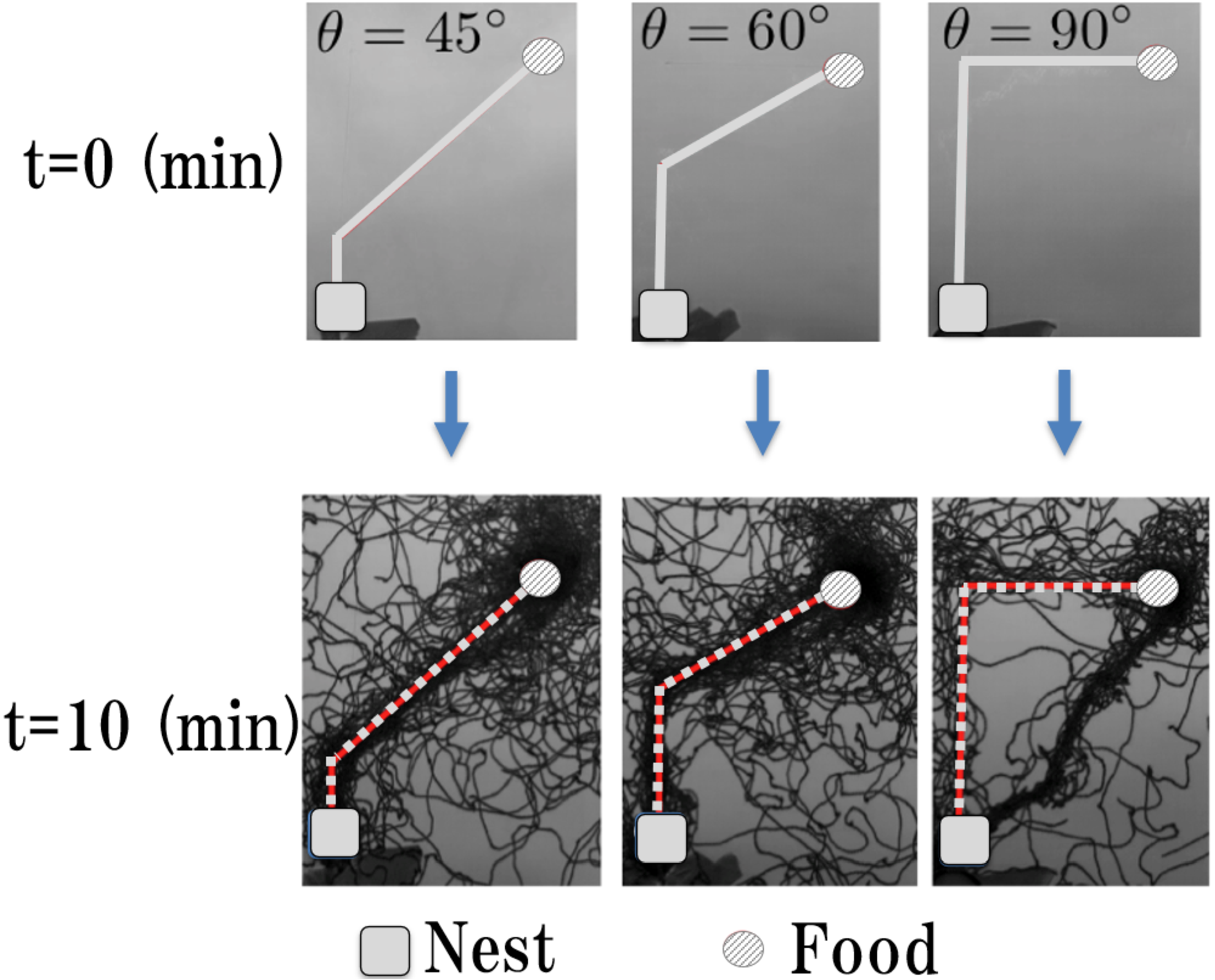}
\end{center}
\caption{Three types of experimental settings 
and results for the observed paths of ants were
reported by the Nishimori group.}
\label{fig:Nishimori2015}
\end{figure}

\begin{description}
\item{1.} \quad
At the beginning of each experiment,
the preliminarily extracted recruit pheromone
was applied along one of the three kinds of polygonal lines
connecting the nest and the food source,
which we call the `pheromone road'.
One folding point was included in the lines
with turning angles 
$\theta \in \{45^{\circ}, 60^{\circ}, 90^{\circ}\}$
(see Fig.\ref{fig:Nishimori2015}).
The preliminarily extracted
pheromone had a sufficiently high density to maintain a
strong attraction of ants until the end of each experiment.

\item{2.} \quad
From another experiment by
the Nishimori group, it is considered that
\textit{Lasius Japonicus} can determine landmarks 
in the order of 10 cm \cite{Shi13}.
Hence, ants in the present setup could
recognize the direction of the nest from the food source.
This direction along the optimal path from the
food source to nest is different from
the direction from the food sources
indicated by the `pheromone road' as mentioned above.
\end{description}

After getting food at the food source, ants start homing
walk, and a conflict between the direction along the
`pheromone road' and the straight direction to the nest
requires ants to make decision on which they rely
chemical cues or visual cues.
The degree $\alpha$ of this conflict becomes larger
as angle $\theta$ increases. 

The results are shown in Fig.~\ref{fig:Nishimori2015}.
In the upper pictures, the initially prepared 
`pheromone roads' are indicated by light-gray lines,
each of them have a folding point with
turning angle $\theta=45^{\circ}, 60^{\circ}$, and
$90^{\circ}$, respectively.
Lower pictures show the trajectories of foraging ants
recorded during the middle period 
(approximately 10 min) of each experiment with a duration
of 60 min. 
Only in the case of $\theta=90^{\circ}$, 
a direct path between the nest and food source
was established, whereas in the other cases,
no direct path was established until the end of the
experiment at time $t=60$ min. 

It was claimed \cite{Nis15} that 
the change of the dominant path from the
trailing paths of the `pheromone road' to the direct path
did not correspond to a continuous geometric 
change of the path. Instead,
the direct path spontaneously emerged 
only in the case where $\theta=90^{\circ}$. 

The observation from the experiments above
suggests that, if we introduce suitable parameters
which control the individual motion of ants,
there exist critical values of these parameters. 
The change in the dominant path can occur 
only within the `supercritical regime' of these parameters.
In this paper, we introduce two types of
\textit{order parameters} that
characterize the global changes in the path patterns
of foraging ants during homing walks from
the food source to nest. 
Roughly speaking, they are given by 
\begin{align}
M &= \frac{\sharp\{\mbox{pioneering ants and their followers}\}}
{\sharp\{\mbox{homing ants}\}},
\nonumber\\
\widetilde{M} &= \frac{\sharp\{\mbox{ants trailing the `pheromone road'}\}}
{\sharp\{\mbox{homing ants}\}},
\label{eqn:order_parameters_0}
\end{align}
where $\sharp\{\mbox{elements}\}$ denotes the total number of elements in 
a set $\{\mbox{elements}\}$.
The precise definitions of $M$ and $\widetilde{M}$ are provided
in Section \ref{sec:homing_duration}.
We also discuss the possibility of considering
global changes in path patterns
as \textit{continuous phase transitions} 
associated with \textit{critical phenomena} 
\cite{CD98,MD99,VZ12}.

\SSC
{Discrete-time Stochastic Model on a Square Lattice}
\label{sec:models}
\subsection{Switching random walks interacting through pheromone field }
\label{sec:algorithm}
\begin{description}
\item{\bf [General setting]} \,
Consider an $(L-1) \times (L-1)$ square region 
on a square lattice.
The set of vertices (sites) is given by
\begin{equation}
V_L:=\{v=(x, y): x \in \{0, 1, \dots, L-1 \},
y \in \{0, 1, \dots, L-1 \} \}.
\label{eqn:sites}
\end{equation}
We denote the Euclidean distance between two vertices
$v_1, v_2 \in V_L$ as $|v_1-v_2|$.

With discrete time
\begin{equation}
t \in \N_0:=\{0,1, \dots\},
\label{eqn:time}
\end{equation}
we consider a set of random walks 
of a given number of particles,
$N \in \N:=\{1, 2, \dots\}$, on $V_L$.
They are interacting with each other through 
the time-dependent pheromone field as explained below.
However, we do not impose any exclusive interactions
between particles. Hence, 
each vertex $v \in V_L$ can be occupied by
any number of particles at each time.

The \textit{nest} and \textit{food sources} are assumed to be
located at the origin $v_{\rm n}:=(0, 0)$ 
and the most upper-right vertex 
$v_{\rm f}:=(L-1, L-1)$, respectively. 

All particles start at $v_{\rm n}$ 
successively from initial time $t=0$
to $t=N-1$ and perform
a drifted random walk toward $v_{\rm f}$. 
After arriving at $v_{\rm f}$, each particle immediately
starts the drifted random walk from $v_{\rm f}$ 
toward $v_{\rm n}$
and then returns to $v_{\rm n}$. 
In this way, 
the particles shuttle between $v_{\rm n}$ and 
$v_{\rm f}$ within a given period $T$.
When a particle is in the time period
for walking from $v_{\rm n}$ to $v_{\rm f}$,
we say that it is in an \textit{exploring period},
and when it is in the time period
for walking from $v_{\rm f}$ to $v_{\rm n}$,
it is said to be in a \textit{homing period}.

We denote two types of hopping rules
that represent the \textit{slow mode} and \textit{fast mode} 
of the drifted random walk, respectively. 
The particles are labeled by $j=1, \dots, N$.
At each time $t \in \N_0$, 
the state of the particle is specified by
\begin{align*}
&\mbox{the location}: \, \, v \in V_L,
\nonumber\\
&\mbox{and the types of hopping : 
slow mode (s) or fast mode (f)}.
\end{align*}
Hence, at each time $t \in \N_0$, 
the $N$-particle configuration 
is given by a set 
\begin{equation}
X(t):=\{X_j(t): j=1, \dots, N\}
\label{eqn:X1}
\end{equation}
of the pairs of random variables,
\begin{equation}
X_j(t)=(v_j(t), \sigma_j(t)),
\quad
v_j(t) \in V_L, \quad
\sigma_j(t) \in \{{\rm s}, {\rm f}\},
\label{eqn:X2}
\end{equation}
$j=1, \dots, N$. 
We define a discrete-time stochastic process, 
$(X(t))_{t \in \N_0}$.

\item{\bf [Time-dependent pheromone field]} \,

The pheromone is placed on the vertices, and its intensity
is expressed by a non-negative integer,
\begin{equation}
f(v, t) \in \N_0,
\quad v \in V_L, \quad t \in \N_0.
\label{eqn:f}
\end{equation}
That is, $f(v, t)=f \in \N_0$ implies that 
there are $f$ units of pheromones on vertex $v \in V_L$
at time $t \in \N_0$.
Define the $\Gamma$-shaped subset of $V_L$,
\begin{align}
\Gamma_{L, \ell}
=& \{ (x, y) \in V_L: 0 \leq x \leq \ell-1, 0 \leq y \leq L-1 \}
\nonumber\\
&
\cup
\{(x,y) \in V_L: 0 \leq x \leq L-1, L-\ell \leq y \leq L-1 \}.
\label{eqn:Gamma1}
\end{align}
We will call $\Gamma_{L, \ell}$ 
the \textit{$\Gamma$-region}, which represents
the `pheromone road'.
The variable $\ell$ indicates the
\textit{width of the $\Gamma$-region}.
See Fig.\ref{fig:paths}. 

\begin{figure}[ht]
\begin{center}
\includegraphics[width=0.4\textwidth]{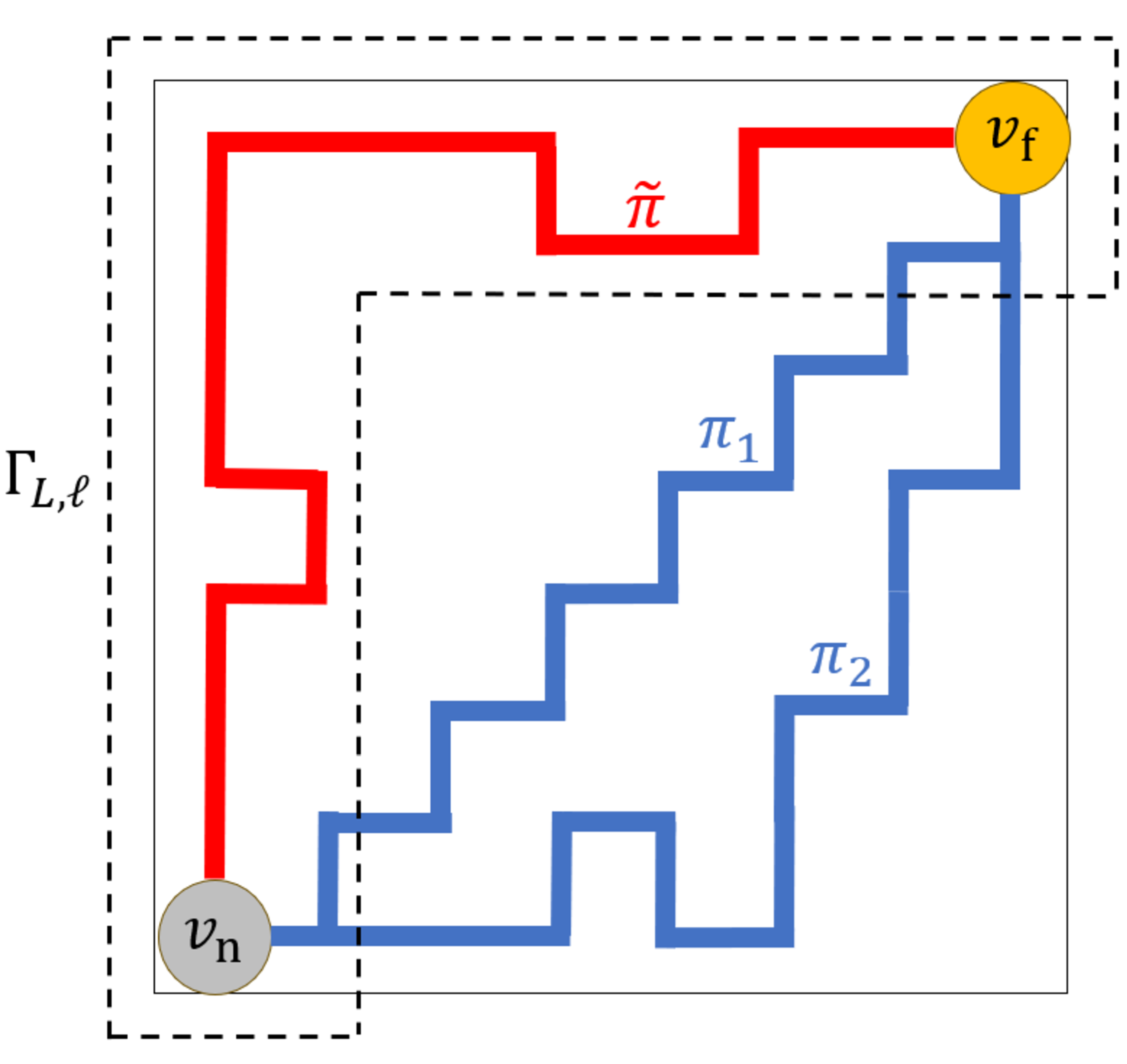}
\end{center}
\caption{The setting of the model in the square region
$V_L$ is illustrated. 
Paths from the food source $v_{\rm f}$
to nest $v_{\rm n}$ during the homing period
are classified. 
The path $\widetilde{\pi}$ shown by the red line is
an example of $\Gamma$-path  
which is completely included in 
the $\Gamma$-region, $\Gamma_{L, \ell}$.
Two examples of nearly optimal paths are 
given by $\pi_1$ and $\pi_2$, which are 
indicated by the blue lines.
Some parts of the paths are outside $\Gamma_{L, \ell}$
and each homing duration $A$ satisfies 
$2(L-1) \leq A \leq 2(L-1)+\bra \Delta A_0 \ket/2$. 
}
\label{fig:paths}
\end{figure}

With parameter $f_0 \in \N$, 
the initial value of the pheromone field is given by
\begin{equation}
f(v, 0)=
\begin{cases}
f_0, &
\mbox{if $v \in \Gamma_{L, \ell}$},
\cr
0, &
\mbox{if $v \in V_L \setminus \Gamma_{L, \ell}$},
\end{cases}
\label{eqn:Gamma3}
\end{equation}
where $V_L \setminus \Gamma_{L, \ell}$ denotes
the complement to $\Gamma_{L, \ell}$ in $V_L$.
With the parameters $f_1, f_2 \in \N_0$,
for each arrival of a particle at each vertex $v \in V_L$, 
the secretion of pheromones by an ant
is represented by the following intensity increment,
\begin{equation}
f \to 
\begin{cases}
f+f_1, & \mbox{in an exploring period},
\cr
f+f_2, & \mbox{in a homing period}.
\end{cases}
\label{eqn:phe1}
\end{equation}

If $v \in V_L \setminus \Gamma_{L, \ell}$, 
each unit of the pheromone evaporates with
probability $q$ independently 
at each time step $t \to t+1$.
In other words, at time step $t \to t+1$, 
\begin{equation}
f \to f-n 
\quad
\mbox{with probability} \quad 
\binom{f}{n} q^n (1-q)^{f-n}, \quad
n=0,1, \dots, f.
\label{eqn:evaporation}
\end{equation}
However, if $v \in \Gamma_{L, \ell}$,
only the excess $f$ over $f_0$ is evaporated.
That is, if and only if $f > f_0$,  
\begin{equation}
f \to f-n 
\quad
\mbox{with probability} \quad 
\binom{f-f_0}{n} q^n (1-q)^{f-f_0-n}, \quad
n=0,1, \dots, f-f_0.
\label{eqn:evaporation2}
\end{equation}

\item{\bf [Hopping rule in an exploring period]} \,
When the ants are in the exploring period,
they are all in the slow mode, 
$\sigma_j(t) \equiv {\rm s}$. 
For $X_j(t)=(v_j(t), {\rm s})$, we define
a set of nearest neighboring vertices of $v_j(t)$ as
\begin{align}
\Lambda^{\rm s}_j(t) &:=
\begin{cases}
\{v \in V_L: |v-v_j(t)|=1\}
=\{(1,0), (0,1)\},
&
\mbox{if $v_j(t)=v_{\rm n}$},
\cr
\{v \in V_L: |v-v_j(t)|=1, v \not= v_j(t-1)\},
&
\mbox{if $v_j(t) \in V_L \setminus \{v_{\rm n}\}$}.
\end{cases}
\label{eqn:Lambda_s}
\end{align}
Then, at each time step $t \to t+1$, 
\begin{equation}
v_j(t) \to v 
\quad \mbox{with probability} \quad
\frac{f(v, t)}
{\sum_{w \in \Lambda^{\rm s}_j(t)} f(w, t)},
\quad \mbox{if and only if $v \in \Lambda^{\rm s}_j(t)$},
\end{equation}
provided $\sum_{w \in \Lambda^{\rm s}_j(t)} f(w, t)>0$.
If $\sum_{w \in \Lambda^{\rm s}_j(t)} f(w, t)=0$;
that is, $f(v, t)=0$ for all $v \in \Lambda^{\rm s}_j(t)$,
then
\begin{equation}
v_j(t) \to v 
\quad \mbox{with probability} \quad
\frac{1}{|\Lambda_j^{\rm s}(t)|},
\quad \mbox{if and only if 
$v \in {\Lambda}^{\rm s}_j(t)$},
\label{eqn:exception1}
\end{equation}
where $|\Lambda_j^{\rm s}(t)|$ is
the total number of vertices in $\Lambda^{\rm s}_j(t)$.
Note that the previous position $v_j(t-1)$ is not included
in $\Lambda^{\rm s}_j(t)$ to prevent backward walking.

\begin{description}
\item{\bf (Prohibition against cyclic motions)}
In an exploring period, if 
$v_j(t+1)-v_j(t) \in \{(-1, 0), (0, -1)\}$, that is, 
the step is leftward or downward, 
hopping is regarded as a \textit{retrograde motion}. 
We prohibit any succession of such retrograde motions
to avoid the cyclic motions of walkers.
When the particle is in the domain $\Gamma_{L, \ell}$,
this prohibition rule is strengthened such that
all downward steps are forbidden if $0 \leq x \leq \ell-1$,
and all leftward steps are forbidden 
if $L-\ell \leq y \leq L-1$.
\end{description}

\item{\bf [Hopping rules
in a homing period]} \,
\begin{description}
\item{\bf (Slow mode)} \,
For $X_j(t)=(v_j(t), {\rm s})$ in a homing period, 
we define
a set of nearest neighboring vertices of $v_j(t)$ as
\begin{align}
\widetilde{\Lambda}^{\rm s}_j(t) &:=
\begin{cases}
\{v \in V_L: |v-v_j(t)|=1, v \not= v_j(t-1)\},
\cr
\hskip 5cm
\mbox{if $v_j(t) \in V_L \setminus \{v_{\rm f} \}$},
\cr
\{v \in V_L: |v-v_j(t)|=1\}=\{(L-2, L-1), (L-1, L-2) \},
\cr
\hskip 5cm
\mbox{if $v_j(t)=v_{\rm f}$}.
\end{cases}
\label{eqn:tilde_Lambda_s}
\end{align}
Note that the previous position $v_j(t-1)$ is not
included in $\widetilde{\Lambda}^{\rm s}_j(t)$ 
to prevent backward walking.
Then, at each time step $t \to t+1$, 
\begin{equation}
v_j(t) \to v 
\quad \mbox{with probability} \quad
\frac{f(v, t)}
{\sum_{w \in \widetilde{\Lambda}^{\rm s}_j(t)} f(w, t)}.
\quad \mbox{if and only if 
$v \in \widetilde{\Lambda}^{\rm s}_j(t)$},
\end{equation}
provided $\sum_{w \in \widetilde{\Lambda}^{\rm s}_j(t)} f(w, t)>0$.
If $\sum_{w \in \widetilde{\Lambda}^{\rm s}_j(t)} f(w, t)=0$;
that is, $f(v, t)=0$ for all $v \in \widetilde{\Lambda}^{\rm s}_j(t)$,
then
\begin{equation}
v_j(t) \to v 
\quad \mbox{with probability} \quad
\frac{1}{|\widetilde{\Lambda}_j^{\rm s}(t)|},
\quad \mbox{if and only if 
$v \in \widetilde{\Lambda}^{\rm s}_j(t)$}.
\label{eqn:exception2}
\end{equation}

\begin{description}
\item{\bf (Prohibition against cyclic motions)}
In a homing period, if 
$v_j(t+1)-v_j(t) \in \{(1, 0), (0, 1)\}$, that is, 
the step is rightward or upward, 
the hopping is regarded as a \textit{retrograde motion}. 
We prohibit any succession of such retrograde motions
to avoid the cyclic motions of walkers. 
When the particle is in the domain $\Gamma_{L, \ell}$,
this prohibition rule is strengthened such that
all rightward steps are forbidden if $L-\ell \leq y \leq L-1$,
and all upward steps are forbidden if $0 \leq x \leq \ell-1$.
\end{description}
\item{\bf (Fast mode)} \,
For $X_j(t)=(v_j(t), {\rm f})$ in a homing period, 
we define
a subset of the nearest neighboring vertices of $v_j(t)$ as
\begin{equation}
\widetilde{\Lambda}^{\rm f}_j(t) 
:=\{ v \in V_L: |v-v_j(t)|=1, |v| < |v_j(t)|\}.
\label{eqn:Lambda_f}
\end{equation}
Then, in the time step $t \to t+1$, 
\begin{equation}
v_j(t) \to v 
\quad \mbox{with probability} \quad
\frac{1}{|\widetilde{\Lambda}_j^{\rm f}(t)|},
\quad \mbox{if and only if 
$v \in \widetilde{\Lambda}^{\rm f}_j(t)$}.
\label{eqn:fastA1}
\end{equation}
Because we have considered the system in
a subset $V_L$ of the square lattice, 
$|\widetilde{\Lambda}^{\rm f}_j(t)| \in \{1, 2\}$. 
If $x=0$ or $y=0$ in $v_j(t)=(x,y)$, then, 
$|\widetilde{\Lambda}^{\rm f}_j(t)|=1$, and hence, 
hopping in fast mode is deterministic. 

\item{\bf (Switching)} \,
We introduce parameters 
$\gamma_{\rm sf}, \gamma_{\rm fs} \in [0, 1]$.
Then, in time step $t \to t+1$ in the homing period, 
we change the mode of hopping as
\begin{align}
\sigma_j(t)&={\rm s} \to {\rm f}
\quad \mbox{with probability $\gamma_{\rm sf}$},
\nonumber\\
\sigma_j(t)&={\rm f} \to {\rm s}
\quad \mbox{with probability $\gamma_{\rm fs}$}. 
\label{eqn:switching}
\end{align}
This implies that the mode does not change with probability
$1-\gamma_{\rm sf}$
(resp.~$1-\gamma_{\rm fs}$), when the particle is 
slow (resp.~fast) modes at each time-point. 

\item{\bf Remark} \,
Consider the extreme case with $\gamma_{\rm sf}=1$
and $\gamma_{\rm fs}=0$; hence, all particles are in 
fast mode in the homing period.
In this case, the path length from $v_{\rm f}$ to
$v_{\rm n}$ is fixed at the minimum value.
Because our model is constructed on
$V_L$ given by \eqref{eqn:sites}, 
the minimal path length is $2(L-1)$.
We notice that the total number of such optimal walks
is $2^{L-1}$ and some of them are completely included
in the $\Gamma$-region representing
the `pheromone road';
an example is a $\Gamma$-shaped walk
along the uppermost and leftmost edges
of $V_L$. These paths are referred to as 
$\Gamma$-paths in Section \ref{sec:Gamma_path}.
This is because of the simple square lattice discretization 
of the space 
which we adopted in our modeling.
In the real experiments described in 
Section \ref{sec:experiment}, 
as a matter of course, the Euclidean length of
direct path between the food source and the nest
is shorter than the length of `pheromone road'
as we can see in the lower picture for 
$\theta=90^{\circ}$ in Fig.~\ref{fig:Nishimori2015}.
Hence, purely visual-cues-mediated walks
would not be realized in any trajectory trailing the
`pheromone road'.
In Section \ref{sec:Gamma_path}, we consider
the opposite extreme case with $\gamma_{\rm sf}=0$
and $\gamma_{\rm fs}=1$.
Because of our simple square lattice discretization of the space,
some of the $\Gamma$-paths can have an optimal length
$2(L-1)$ as mentioned above;
however, they were not observed in our simulations.
All the observed $\Gamma$-paths are far from optimal.
This phenomenon is caused by our setting so that
the width of $\Gamma$-region,
that is the number of `lanes' in the $\Gamma$-region,
is $\ell \geq 2$ and
`lane changes' are enhanced by the pheromone field,
which make the $\Gamma$-paths include many 
redundant walk.
In conclusion, simple square lattice distretization 
of the space
allows us to simulate the transitions between
the redundant $\Gamma$-paths and 
nearly optimal paths depending on
the switching parameters $\gamma_{\rm sf}$
and $\gamma_{\rm fs}$,
which are expected from the experiments
by the Nishimori group. 
\end{description}

\item{\bf [Initial condition, update rule,  random switching at food source,
and time duration]} 
\begin{description}
\item{(i)} \,
We consider an initial configuration such that
all $N$ particles are in the slow mode and 
start successively from the origin,
\begin{equation}
X_j(j-1)=(0, {\rm s}),
\quad j=1, \dots, N.
\label{eqn:initial1}
\end{equation}

\item{(ii)} \,
We perform hopping of the particles and
switch the mode in homing period
sequentially according to the numbering of 
particles $j=1, \dots, N$. 
The pheromone field is updated at each time step. 

\item{(iii)} \,
When $v_j(t)=v_{\rm f}$, that is, an ant arrives at 
the food source, switching between
the slow and fast modes occurs randomly: 
\begin{equation}
\sigma_j(t+1)=
\begin{cases}
{\rm s}, & \mbox{with probability 1/2},
\cr
{\rm f}, & \mbox{with probability 1/2}.
\end{cases}
\label{eqn:random_switching}
\end{equation}

\item{(iv)} \,
Let $\tau(n)$ be the time duration such that 
the total number of particles which have
returned to the nest $v_{\rm n}$ 
is $n$.
We perform each numerical simulation up to the time step
\begin{equation}
T=\tau(\kappa N),
\label{eqn:T}
\end{equation}
where $\kappa$ denotes a parameter. 
That is, during $T$, totally $\kappa N$ particles shuttle 
between the nest and food source. 
\end{description}
\end{description}

\subsection{Parameter setting}
\label{sec:settings}

We set the parameters concerning 
the time-dependent pheromone field as 
\begin{align}
&f_0=1000, 
\quad f_1=0, \quad f_2=10,
\quad q=0.01.
\label{eqn:setting1}
\end{align}
We assume that 
the density of the particles is fixed as
\begin{equation}
\rho:=\frac{N}{L^2}=\frac{1}{8}=0.125, 
\label{eqn:setting1b}
\end{equation}
the ratio of the width 
$\ell$ of the $\Gamma$-region defined by
\eqref{eqn:Gamma1} 
with respect to $L$ is given by
\begin{equation}
\frac{\ell}{L}=\frac{1}{20},
\label{eqn:setting1c}
\end{equation}
and each simulation is performed for the time duration
\eqref{eqn:T} with
\begin{equation}
\kappa=4.
\label{eqn:setting1d}
\end{equation}

We have performed numerical simulations
of the proposed stochastic model on
square lattices of sizes $L=40$, 
$60$, and $80$. In the present paper, 
we report the dependence of 
the simulation results on the
switching parameters $\gamma_{\rm sf}$ and
$\gamma_{\rm fs}$ which are changed in 
the interval $[0,1]$. 
We have confirmed that the numerical results given below
do not change qualitatively by changing the 
parameter settings \eqref{eqn:setting1}--\eqref{eqn:setting1d}. 
In particular, setting \eqref{eqn:setting1d} for
the time duration of each simulation is sufficient to
observe stable distributions of the homing duration,
which are reported in Section \ref{sec:homing_duration}.
The validity of the $L$-dependence of the parameters
\eqref{eqn:setting1b} and \eqref{eqn:setting1c} is 
discussed in Section \ref{sec:data_collapse}.
A systematic study of the quantitative dependence of the model
on parameters other than $L, \ell, N$ and $\gamma_{\rm sf}$
(with the relationship $\gamma_{\rm fs}=1-\gamma_{\rm sf}$) 
is considered as one of the future problems discussed
in item (v) in Section \ref{sec:future}.
Table \ref{table:parameters} summarizes
the setting, changing, and fixed parameters of 
the present model.

\begin{table}[hbtp]
  \caption{List of the setting, changing, and fixed parameters
  in the present discrete-time stochastic model on a square lattice}
  \label{table:parameters}
  \centering
  \begin{small}
  \begin{tabular}{c|l|l}
    \hline
    \hline
    setting parameters  &   &  setting  \\
    \hline \hline
    $L$  & system size  & 40, 60, 80 \\
    \hline
    $\ell$ & width of the $\Gamma$-region &$L/20$ \\
    \hline
    $N$ & number of particles & $\rho L^2=L^2/8$ \\
    \hline
    \hline
    changing parameters & & range of value \\
    \hline
    \hline
    $\gamma_{\rm sf}$  
       &switching probability from slow mode to fast mode
       & $[0,1]$ \\
    \hline
    $\gamma_{\rm fs}$  
       &switching probability from fast mode to slow mode
       & $[0,1]$ or $1-\gamma_{\rm sf}$ \\
    \hline
    \hline
    fixed parameters &  & value \\
    \hline
    \hline
    $f_0$ & initial intensity of pheromone in the $\Gamma$-region
           & 1000 \\
    \hline
    $f_1$ & amount of pheromone dropped by 
    an exploring particle       
          & 0 \\
    \hline
    $f_2$ & amount of pheromone dropped 
    by a homing particle
          & 10 \\
    \hline
    $q$ & evaporation rate of pheromone & 0.01 \\
    \hline
    $\kappa$ & time duration of simulation
    $T=\tau(\kappa N)$ & 4 \\
    \hline      
  \end{tabular}
  \end{small}
\end{table}

\SSC
{Distributions of Homing Duration and Order Parameters}
\label{sec:homing_duration}
\subsection{$\Gamma$-paths and
order parameter $\widetilde{M}$}
\label{sec:Gamma_path}
\begin{figure}[ht]
\begin{center}
\includegraphics[width=0.6\textwidth]{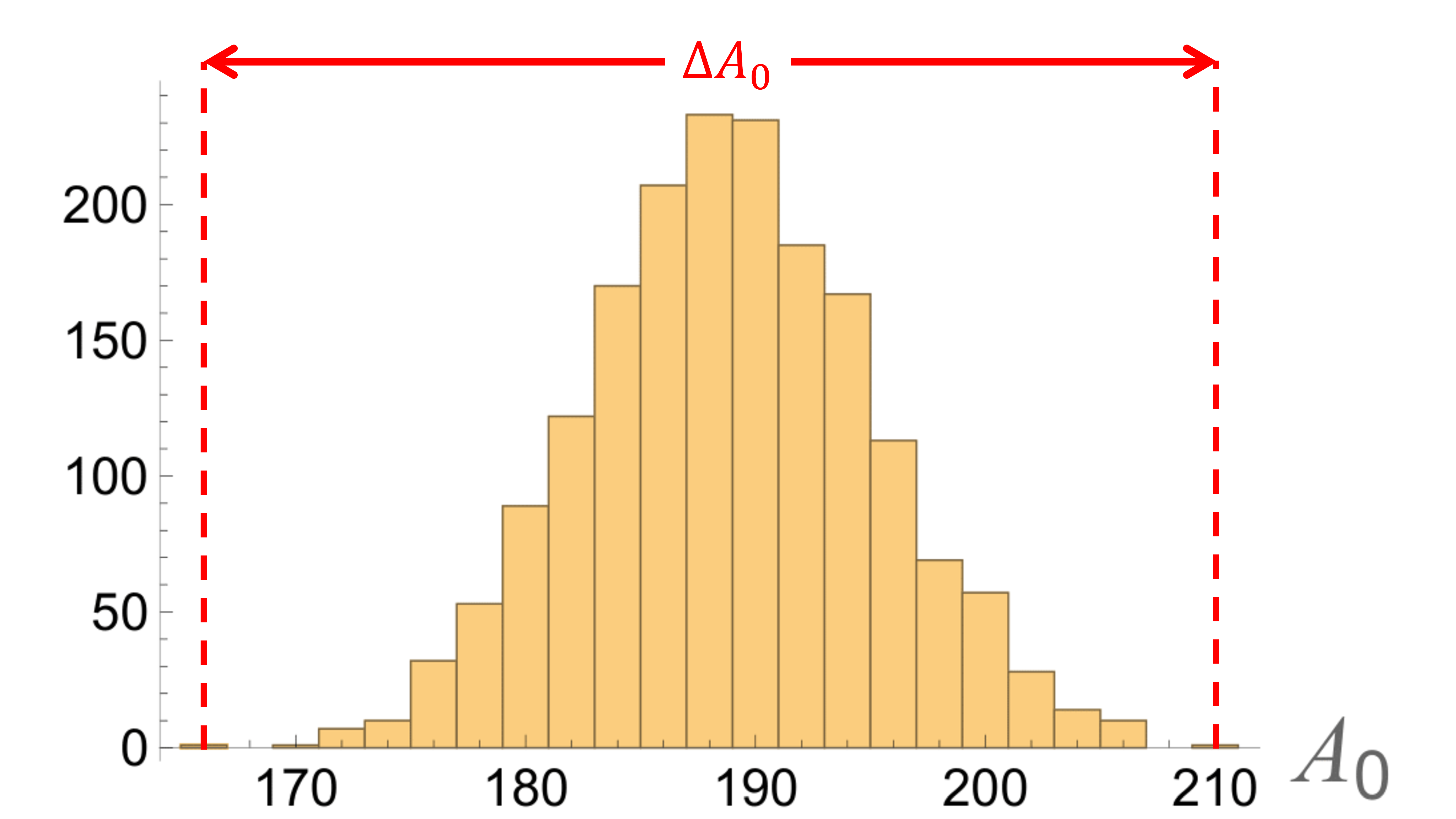}
\end{center}
\caption{
Histogram of the homing duration $A_0$ of particles
in one simulation sample, when
$(\gamma_{\rm sf}, \gamma_{\rm fs})=(0, 1)$, 
in a system of size $L=60$. 
All particles are in slow mode and
trail $\Gamma$-paths.
In this observed distribution,
the mean of $A_0$ is $188.9$, the maximum and minimum 
values are 210 and 166, respectively, and hence, 
$\Delta A_0=44$.
}
\label{fig:G_path_distribution}
\end{figure}

For each particle in the homing period, we trace the path
from the food source $v_{\rm f}$ to nest $v_{\rm n}$,
\begin{equation}
\pi = \{ v_{\rm f}=v_0 \to v_1 \to v_2 \to \cdots
\to v_A = v_{\rm n}\}.
\label{eqn:path1}
\end{equation}
Here, the number of steps in $\pi$ 
indicating the path length is expressed by $A$.
In this study, we refer to the random variable $A$
as the \textit{homing duration}. 

\begin{figure}[ht]
\begin{center}
\includegraphics[width=0.6\textwidth]{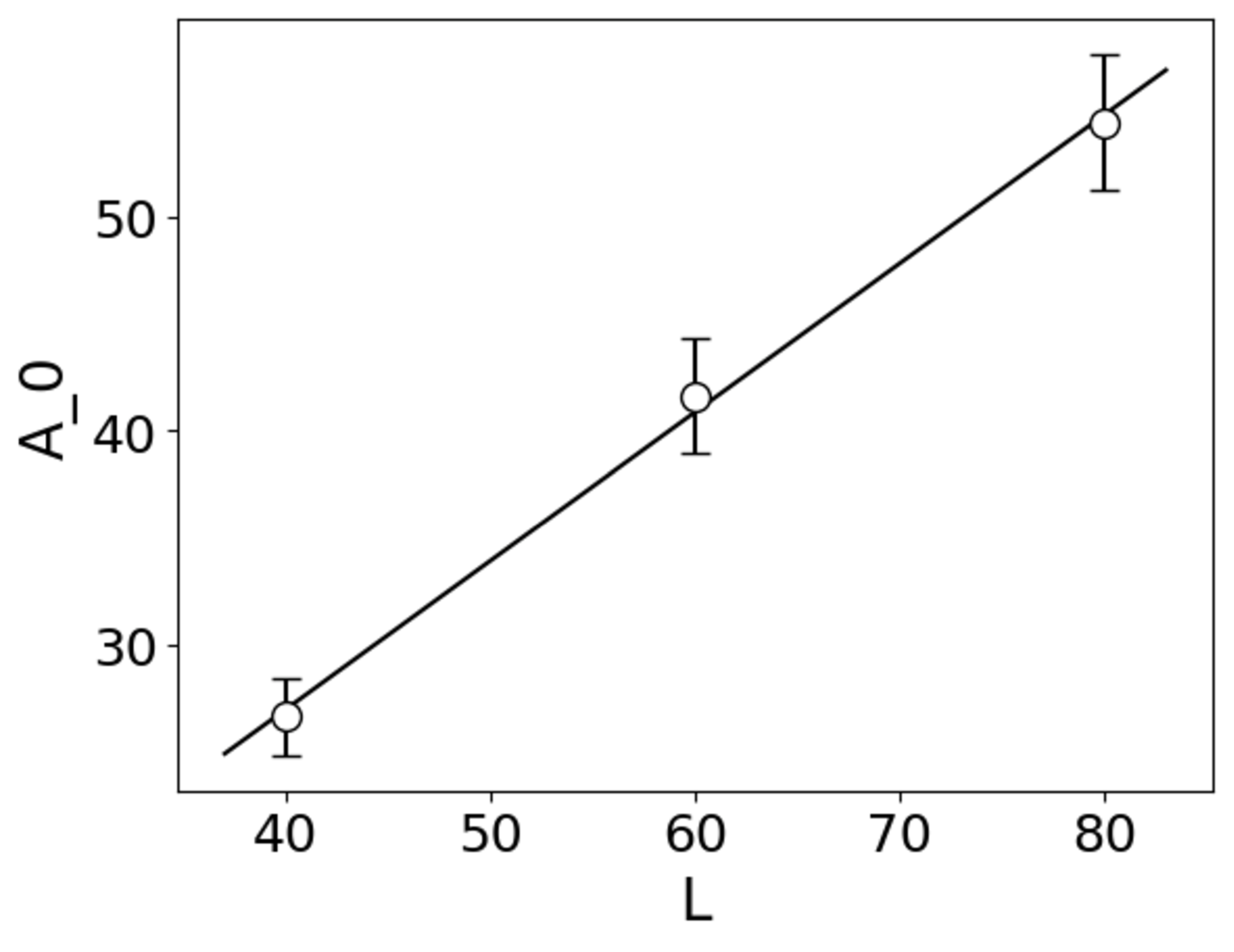}
\end{center}
\caption{
The dependence of $\bra \Delta A_0 \ket$
on $L$ is well described by a linear fitting; 
$\bra \Delta A_0 \ket \simeq 0.7 L-0.83$.
The errors are given by
standard deviations over ten samples of 
numerical simulation for systems of sizes $L=40$ and $60$, 
and five samples for the system of size $L=80$.
}
\label{fig:LvsA0}
\end{figure}

First, we consider the extreme case with
$\gamma_{\rm sf}=0$ and $\gamma_{\rm fs}=1$.
In this case all particles are in the slow mode
following the `pheromone road'.
Hence, all the paths in the homing period are
completely included in the $\Gamma$-region, 
$\Gamma_{L, \ell}$, as defined by \eqref{eqn:Gamma1}.
Generally, if the path in the homing period is completely included in the
$\Gamma$-region, we say that the path is a
\textit{$\Gamma$-path}. 
(See Fig.\ref{fig:paths}.)
In this extreme case, we write the random variable
of the homing duration as $A_0$ by putting the subscript 0.
Figure \ref{fig:G_path_distribution} shows a typical distribution
of the homing duration $A_0$ in one simulation sample
for a system of size $L=60$.
We note that the minimum value of
the homing duration is $2(L-1)=118$, 
which is realized, for example, by 
a $\Gamma$-shaped homing walk along the
uppermost edges and then the leftmost edges 
of $V_L$. 
However, the observed distribution of $A_0$ in this simulation
centered at a larger
value $\simeq 189$.
This implies that slow mode paths are far from optimal
and include many redundant walks 
which are enhanced by the pheromone field.
In the observed distribution, 
we measured width 
$\Delta A_0$ defined as the difference between the
observed maximum and minimum values of $A_0$. 
We calculated the average value
$\bra \Delta A_0 \ket$ for several 
numerical-simulation samples.
The results are as follows:
\begin{align}
\bra \Delta A_0 \ket &= 26.6 \pm 1.8, \quad \mbox{for $L=40$}, 
\nonumber\\
\bra \Delta A_0 \ket &= 41.6 \pm 2.7, \quad \mbox{for $L=60$}, 
\nonumber\\
\bra \Delta A_0 \ket &= 54.4 \pm 3.2, \quad \mbox{for $L=80$},
\label{eqn:DeltaA}
\end{align}
where the errors are given by
the standard deviations over ten samples of 
numerical simulations for systems of sizes $L=40$ and $60$, 
and five samples for the system of size $L=80$.
The above results show a linear dependence of
$\bra \Delta A_0 \ket$ on $L$,
\begin{equation}
\bra \Delta A_0 \ket \simeq 0.7 L - 0.83,
\label{eqn:linear_fitting}
\end{equation}
as shown by 
Fig.\ref{fig:LvsA0}.

We define $\widetilde{M}$ as 
the ratio of the number of observed 
particles following $\Gamma$-paths
to the total number of homing particles in the simulation;
{\it i.e.} $\kappa N = \kappa L^2/8$
;
\begin{equation}
\widetilde{M} :=
\frac{\sharp\{\mbox{homing particle following a $\Gamma$-path}\}}
{\kappa N}.
\label{eqn:M_tilde}
\end{equation}
As previously mentioned, 
$\widetilde{M}=1$, if $(\gamma_{\rm sf}, \gamma_{\rm fs})=(0,1)$.

From now on, we assume the relationship
\begin{equation}
\gamma_{\rm sf}+\gamma_{\rm fs}=1.
\label{eqn:relation1}
\end{equation}
As $\gamma_{\rm sf}$ increases from 0,
$\widetilde{M}$ decreases from 1,
since particles in the fast mode appear
and they make deviation from the `pheromone road'.
In other words, $\widetilde{M}$ plays the role
of the \textit{order parameter} for
the path selection which persists the original
`pheromone road' in the homing period.

\subsection{Nearly optimal paths and
order parameter $M$}
\label{sec:OP_M}

\begin{figure}[htbp]
\begin{center}
  \begin{minipage}[b]{0.65\linewidth}
    \centering
    \includegraphics[keepaspectratio, scale=0.2]{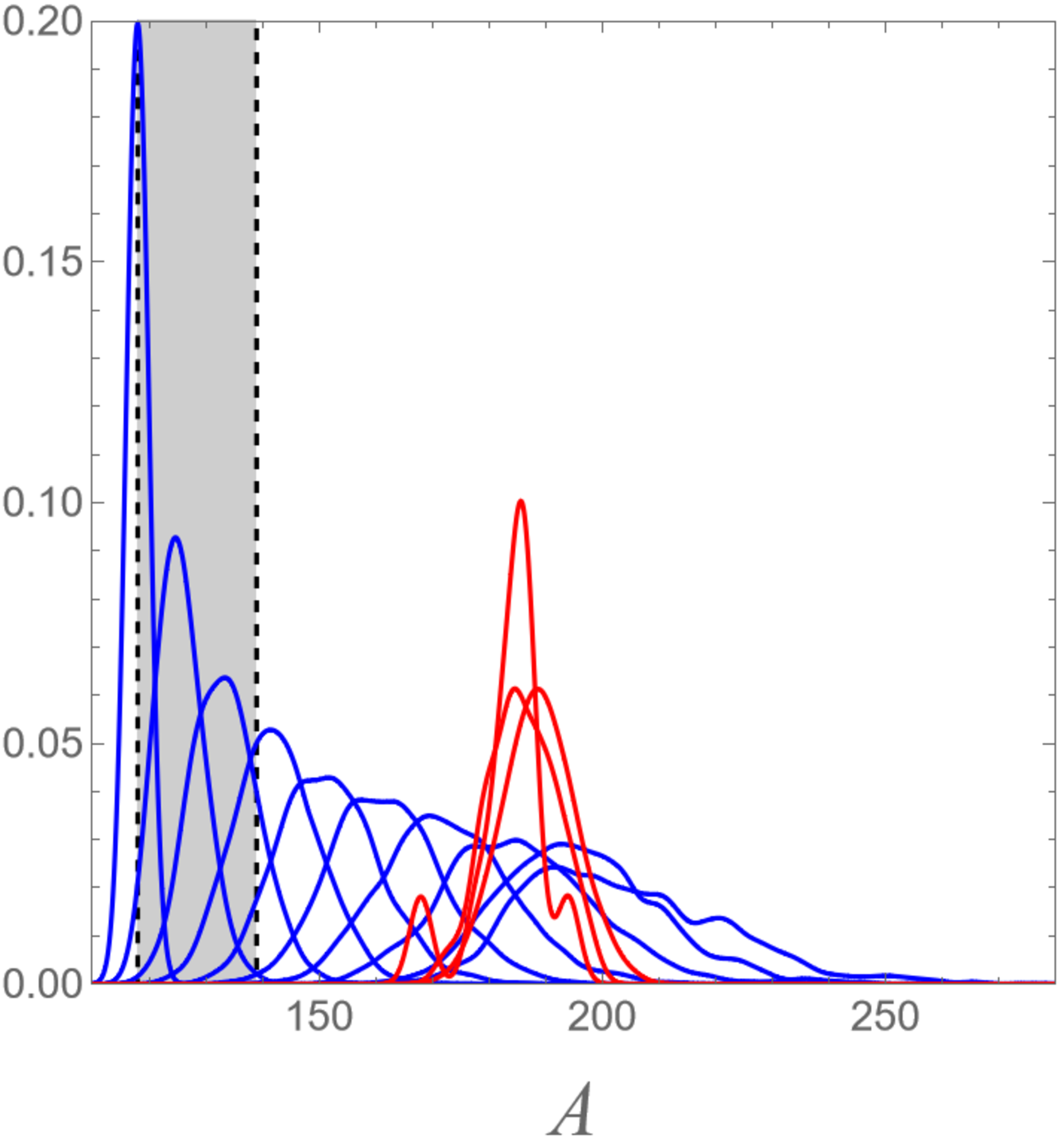}
    \subcaption{Dependence on $\gamma_{\rm sf}$ with the relationship 
$\gamma_{\rm fs}=1-\gamma_{\rm sf}$ of the
distributions of homing duration $A$ of 
$\Gamma$-paths (shown by red curves)
and of non-$\Gamma$-paths (by blue curves)
for the system with $L=60$. 
Three distributions of $A$ are shown for the
$\Gamma$-paths with $\gamma_{\rm sf} =0$, $0.1$, and $0.2$,
and ten for the non-$\Gamma$-paths with
$\gamma_{\rm sf}=0.1$, $0.2$, $0.3$, $0.4$, $0.5$, $0.6$, $0.7$, $0.8$, $0.9$,
and $1$.
In both the cases, 
the right edge (the point at which the right tail of
the distribution becomes zero) decreases monotonically
as $\gamma_{\rm sf}$ increases.}
  \end{minipage}
  \\
\vskip 0.3cm
  \begin{minipage}[b]{0.65\linewidth}
    \centering
    \includegraphics[keepaspectratio, scale=0.2]{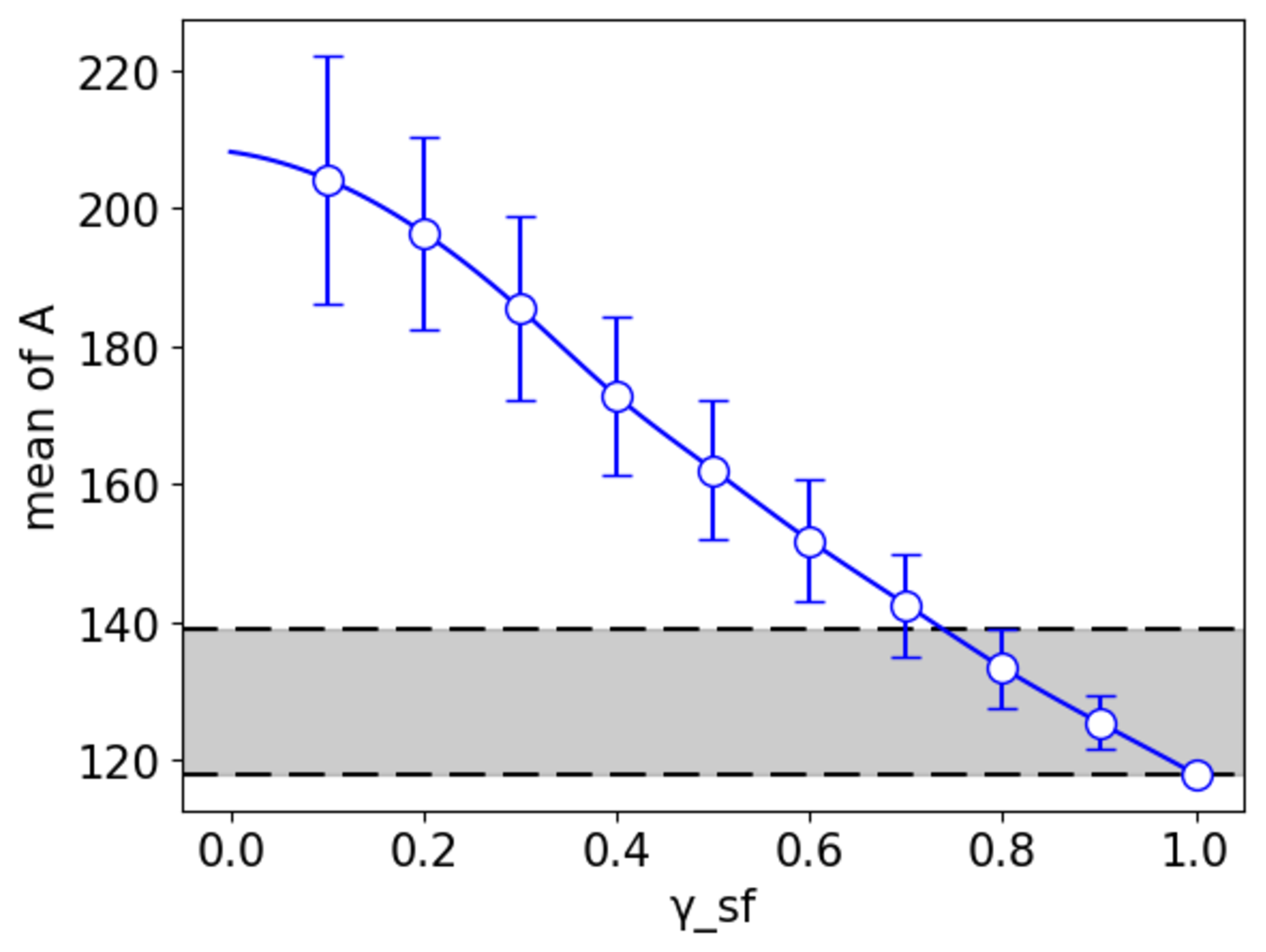}
\subcaption{The means of the distribution of $A$ for 
non-$\Gamma$-paths 
(shown by the blue curves in (a)) are
plotted against $\gamma_{\rm sf} \in (0, 1]$.
The standard deviations of the distribution of $A$ are indicated by
error bars.
}
\end{minipage}
  \caption{Distributions of the homing duration $A$.
The shaded region indicates the range of $A$,
$2(L-1) \leq A \leq 2(L-1)+\bra \Delta A_0 \ket/2$,
for the nearly optimal paths.
For $L=60$, $2(L-1)=118$ and
$\bra \Delta A_0 \ket/2=20.8$ by \eqref{eqn:DeltaA}.}
\label{fig:NGPdist}
\end{center}
\end{figure}
  
For a system of size $L=60$, Fig.~\ref{fig:NGPdist} (a) shows the
dependence of the distribution of the homing duration $A$ 
on $\gamma_{\rm sf}$ with relation \eqref{eqn:relation1}. 
When $\gamma_{\rm sf} \gtrsim 0.3$, $\Gamma$-path
is no longer observed 
and the observed paths are dominated by
\textit{non-$\Gamma$-paths}. 
As $\gamma_{\rm sf}$ increases, 
the observed support of the distribution 
of the homing duration for
the non-$\Gamma$-paths shifts to lower value
region of $A$. 
This means that as $\gamma_{\rm sf}$ increases, 
more pioneer walkers appear to
build shorter paths from the food source to nest 
because particles can be in the fast mode more
frequently. 
In addition to the left shift,
Fig.~\ref{fig:NGPdist} (a) shows that
the bell-shaped distribution
of $A$ for the non-$\Gamma$-paths
becomes sharper as $\gamma_{\rm sf}$ increases.
The monotonic decrease in the mean $\bra A \ket$
and standard deviation $\sigma$ of the distribution of $A$
with the increment of $\gamma_{\rm sf}$ 
is shown in Fig.~\ref{fig:NGPdist} (b) for the non-$\Gamma$-paths. 
This implies that shorter paths are
strengthened by the followers of pioneer walkers
via interactions through the pheromone field. 
As $\gamma_{\rm sf}$ approaches one, 
the distribution quickly sharpens and
appears to condense to 
the shortest value of $A$, \textit{i.e.} 
the length of the optimal path, $2(L-1)=118$. 
In other words, as $\gamma_{\rm sf}$ approaches 1, 
a large number of particles tend to trail through
paths which are very close to the optimal path.
We want to call such paths nearly optimal paths.

To define nearly optimal paths,
we use the value $\bra \Delta A_0 \ket$ determined by
the homing duration distribution when
$(\gamma_{\rm sf}, \gamma_{\rm fs})=(0, 1)$
in Section \ref{sec:Gamma_path}.
If the homing duration $A$ of the path satisfies
the condition
\begin{equation}
2(L-1) \leq A \leq
2(L-1) + \frac{1}{2}\bra \Delta A_0 \ket,
\label{eqn:near_optimal}
\end{equation}
then we say that it is a \textit{nearly optimal path}.
The range \eqref{eqn:near_optimal} of $A$ 
is indicated by the shaded regions in Figs.~\ref{fig:NGPdist} (a) and (b).
Then, we define another order parameter $M$
in addition to $\widetilde{M}$ 
as the ratio of the number of particles following
nearly optimal paths in a homing period
to the total number of homing particles in the simulation,
$\kappa N$, 
\begin{equation}
M :=
\frac{\sharp\{\mbox{homing particle following 
a nearly optimal path}\}}
{\kappa N}.
\label{eqn:M}
\end{equation}
Figure \ref{fig:paths} illustrates an example of $\Gamma$-path
which contributes to the order parameter $\widetilde{M}$
and two examples of nearly optimal paths
which contribute to the order parameter $M$.
We will compare the dependence on $\gamma_{\rm sf}$
of these two types of 
order parameters $M$ and $\widetilde{M}$
varying the system size $L$.
The choice of $\bra \Delta A_0 \ket/2$ as the width
in the condition \eqref{eqn:near_optimal} which defines 
the nearly optimal path is 
to make the dependence of $M$ on $L$ 
comparable with that of $\widetilde{M}$. 
The definitions \eqref{eqn:near_optimal} 
and \eqref{eqn:M} work well, as shown below.

\SSC
{Phase Transitions and Critical Phenomena in Path Selections}
\label{sec:phase_transition}
\subsection{Almost data-collapse 
with respect to system size}
\label{sec:data_collapse}
\begin{figure}[htbp]
\begin{center}
  \begin{minipage}[b]{0.50\linewidth}
    \centering
    \includegraphics[keepaspectratio, scale=0.2]{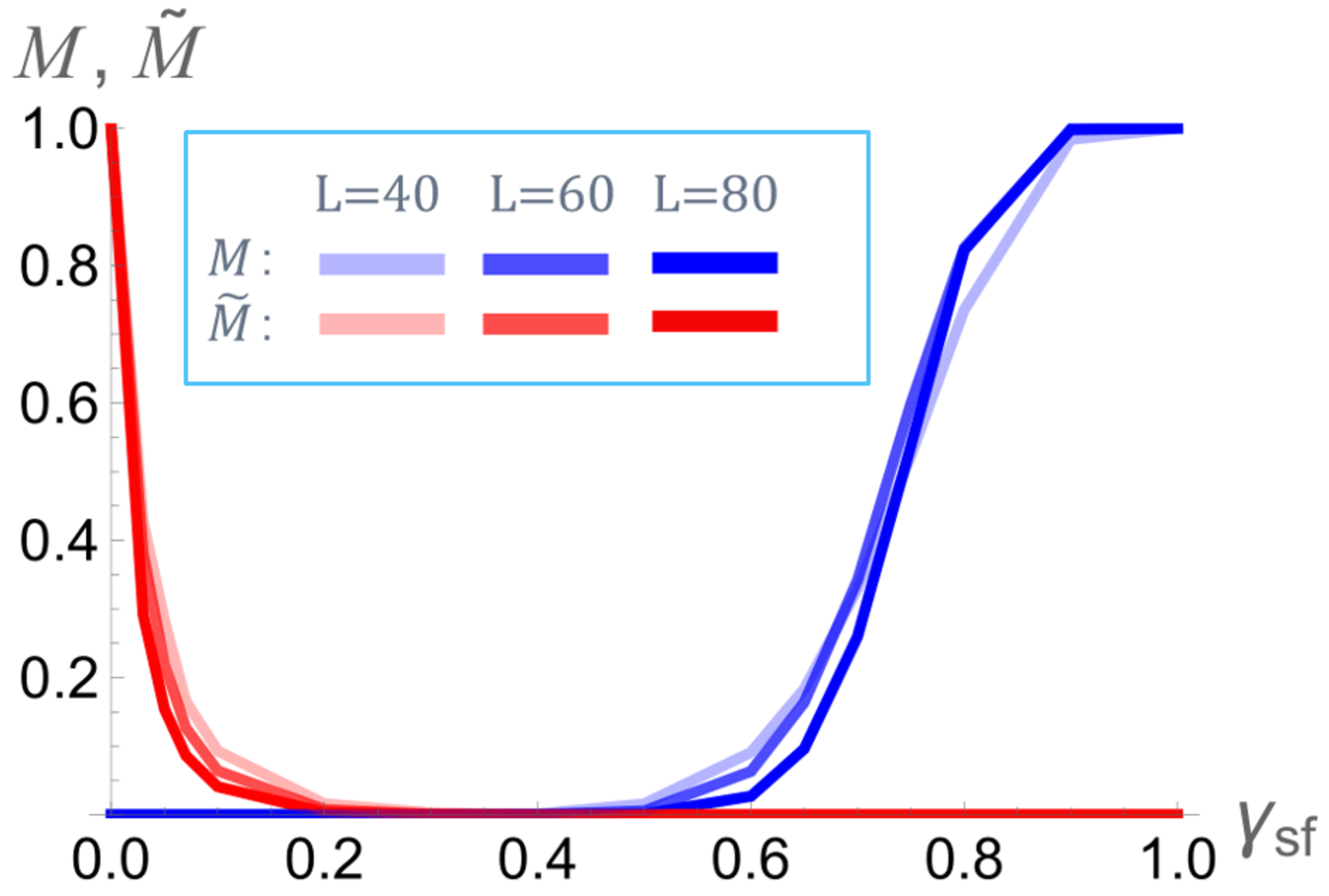}
    \subcaption{
    For each value of $\gamma_{\rm sf}$, 
    we put $\gamma_{\rm fs}=1-\gamma_{\rm sf}$
    and numerically evaluated the order parameters
    $M_L$ and $\widetilde{M}_L$ for the three systems
    of sizes $L=40$, $60$, and $80$, 
    as explained in Section \ref{sec:data_collapse}. 
    A total of $2 \times 3=6$ curves are plotted
    versus $\gamma_{\rm sf}$ in different colors.
    }
  \end{minipage}
  \\
\vskip 0.3cm
  \begin{minipage}[b]{0.50\linewidth}
    \centering
    \includegraphics[keepaspectratio, scale=0.2]{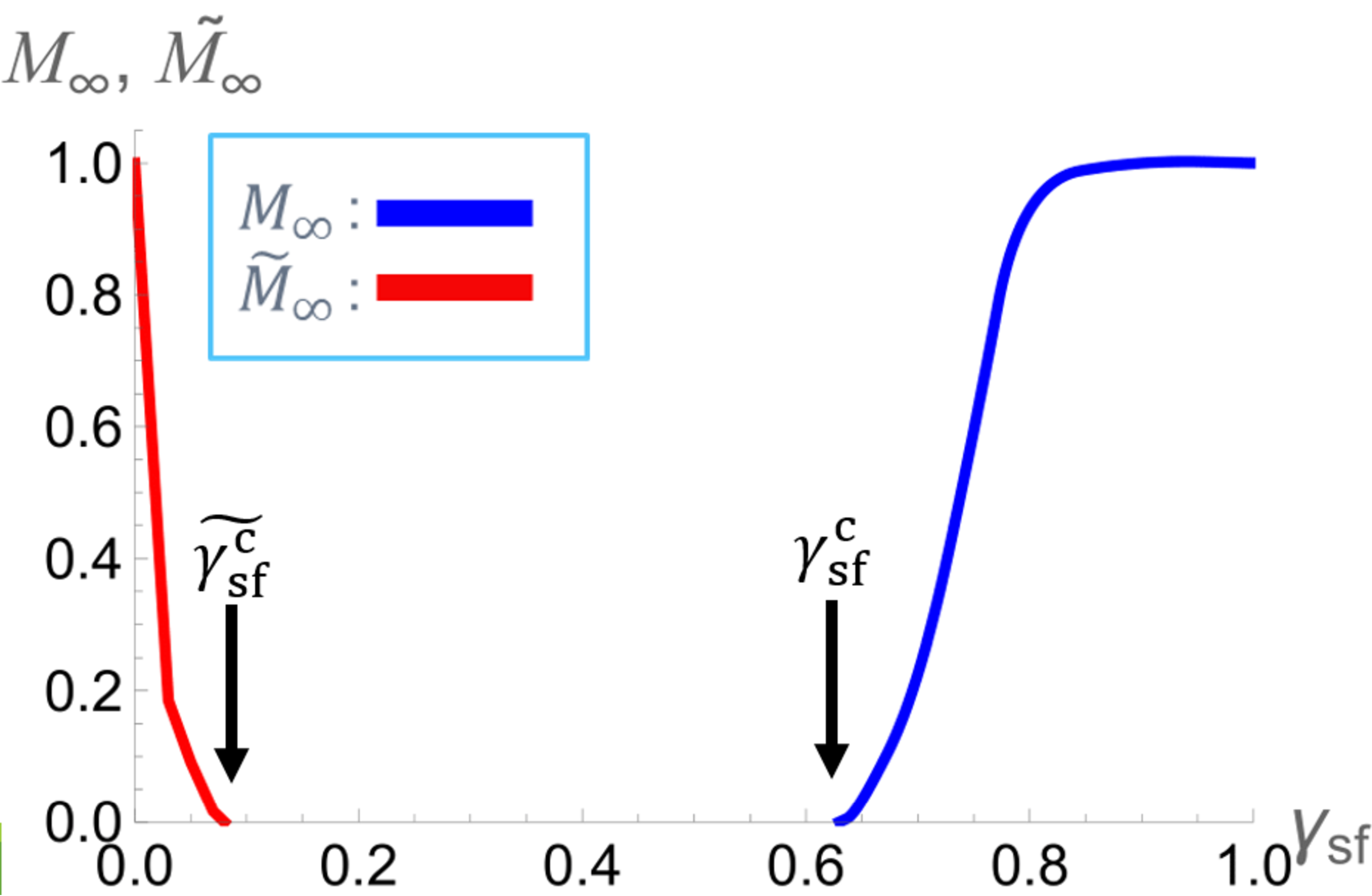}
\subcaption{$M_{\infty}$ and $\widetilde{M}_{\infty}$
for an infinite system are
estimated using $1/L$-plots, 
as described in Section \ref{sec:critical_exponent}.
The critical values $\gamma_{\rm sf}^{\rm c} \simeq 0.63$ 
and $\widetilde{\gamma_{\rm sf}^{\rm c}} \simeq 0.08$ 
are indicated by the arrows.}
\end{minipage}
  \caption{The order parameters $M$ and $\widetilde{M}$
are plotted against $\gamma_{\rm sf} \in [0, 1]$,
where $\gamma_{\rm fs}=1-\gamma_{\rm sf}$.}
\label{fig:OrderParameters}
\end{center}
\end{figure}

Using the relation between the parameters
\eqref{eqn:relation1}, 
Fig.\ref{fig:OrderParameters} (a) shows that
the $\gamma_{\rm sf}$-dependence
of the two types of order parameters,
$M$ and $\widetilde{M}$, 
for systems of different sizes, $L=40$, $60$, and $80$.
For systems of sizes, $L=40$ and $60$, 
(resp. $80$) 
we have obtained 10 samples (resp. 3 samples) 
for each parameter $\gamma_{\rm sf} \in [0, 1]$.
These lines interpolate the average values of the samples.
We have confirmed that the standard deviations are very small,
\textit{e.g.} $3.0 \times 10^{-3}$ for $M$ at $\gamma_{\rm sf}=0.6$
and $1.5 \times 10^{-2}$ for $\widetilde{M}$ at $\gamma_{\rm sf}=0.1$
for $L=60$. Hence, the error bars are omitted, as shown
in Fig.\ref{fig:OrderParameters} (a). 
By definition, $\widetilde{M} \to 1$ as 
$\gamma_{\rm sf} \to 0$, and
$M \to 1$ as $\gamma_{\rm sf} \to 1$. 
We observe that 
when $\gamma_{\rm sf}$ is small, 
$M = 0$ and $\widetilde{M} >0$,
and when $\gamma_{\rm sf}$ is large, 
$M >0$ and $\widetilde{M} = 0$. 
The rapid decrease in $\widetilde{M}$
around $\gamma_{\rm sf} \simeq 0.1$, 
and the rapid
increase in $M$ around $\gamma_{\rm sf} \simeq 0.6$
indicate the global switching phenomena
of path selection between
the $\Gamma$-paths and nearly optimal paths.

Figure \ref{fig:OrderParameters} (a) shows 
that the differences in the values of
$M$ and $\widetilde{M}$ caused by increasing the
system size $L$ are relatively small. 
Roughly speaking, 
this fact is regarded as a kind of
\textit{data collapse}
that shows the independence of 
global switching phenomena 
in path selections 
on size $L$ in the proposed model.
That is, the settings for the 
number of particles, 
$N=\rho L^2 \propto L^2$, as given by \eqref{eqn:setting1b}, 
the width of the $\Gamma$-region
(`pheromone road'), 
$\ell=L/20 \propto L$, given by \eqref{eqn:setting1c}, 
and the condition \eqref{eqn:near_optimal} 
for nearly optimal paths using the
quantity, $\bra \Delta A_0 \ket \simeq a L + b$, 
(see \eqref{eqn:linear_fitting})
are simple, but they
provide a \textit{proper scaling property} 
of the system with respect to
the change in the system size $L$.

\subsection{Evaluations of critical values
and critical exponents}
\label{sec:critical_exponent}

In contrast, 
in the vicinity of the values of $\gamma_{\rm sf}$
at which $\widetilde{M}$ approaches zero
($\gamma_{\rm sf} \simeq 0.1$)
and $M$ emerges ($\gamma_{\rm sf} \simeq 0.6$), 
the changes show systematic dependence on $L$. 
This observation suggests that we will have
\textit{critical phenomena} in the $L \to \infty$
limit and that the present global switching phenomena
in path selection can be considered as
\textit{continuous phase transitions}. 

To explain our method 
for estimating the order parameters
in the $L \to \infty$ limit, we write the
order parameters as $M_L$ and $\widetilde{M}_L$
for a system of size $L$.
At each value of $\gamma_{\rm sf} \in [0,1]$
with \eqref{eqn:relation1}
we assume the following simple dependence of the
order parameters on system size $L$,
\begin{equation}
M_{\infty}=M_L + \frac{c_1}{L},
\quad
\widetilde{M}_{\infty} 
=\widetilde{M}_L+\frac{\widetilde{c_1}}{L},
\quad \mbox{as $L\to \infty$},
\label{eqn:1/L_1}
\end{equation}
where $c_1$ and $\widetilde{c_1}$ are fitting parameters, and
$M_{\infty}$ and $\widetilde{M}_{\infty}$ are the estimated
values of the 
order parameters of an infinite system.
(See item (ii) in Section \ref{sec:future}
for a possible improvement by introducing
the critical exponents $\nu$ and $\widetilde{\nu}$.) 
By plotting the values of $M_L$ (resp. $\widetilde{M}_L$)
with respect to $1/L$, the $L \to \infty$ limit
of the order parameter $M_{\infty}$ 
(resp. $\widetilde{M}_{\infty}$)
is evaluated using the `$y$-intercept'.
The estimated curves of $M_{\infty}$ 
and $\widetilde{M}_{\infty}$ are shown 
for $\gamma_{\rm sf} \in [0,1]$ 
in Fig.~\ref{fig:OrderParameters}~(b).

\begin{figure}[htbp]
\begin{center}
  \begin{minipage}[b]{0.50\linewidth}
    \centering
    \includegraphics[keepaspectratio, scale=0.3]{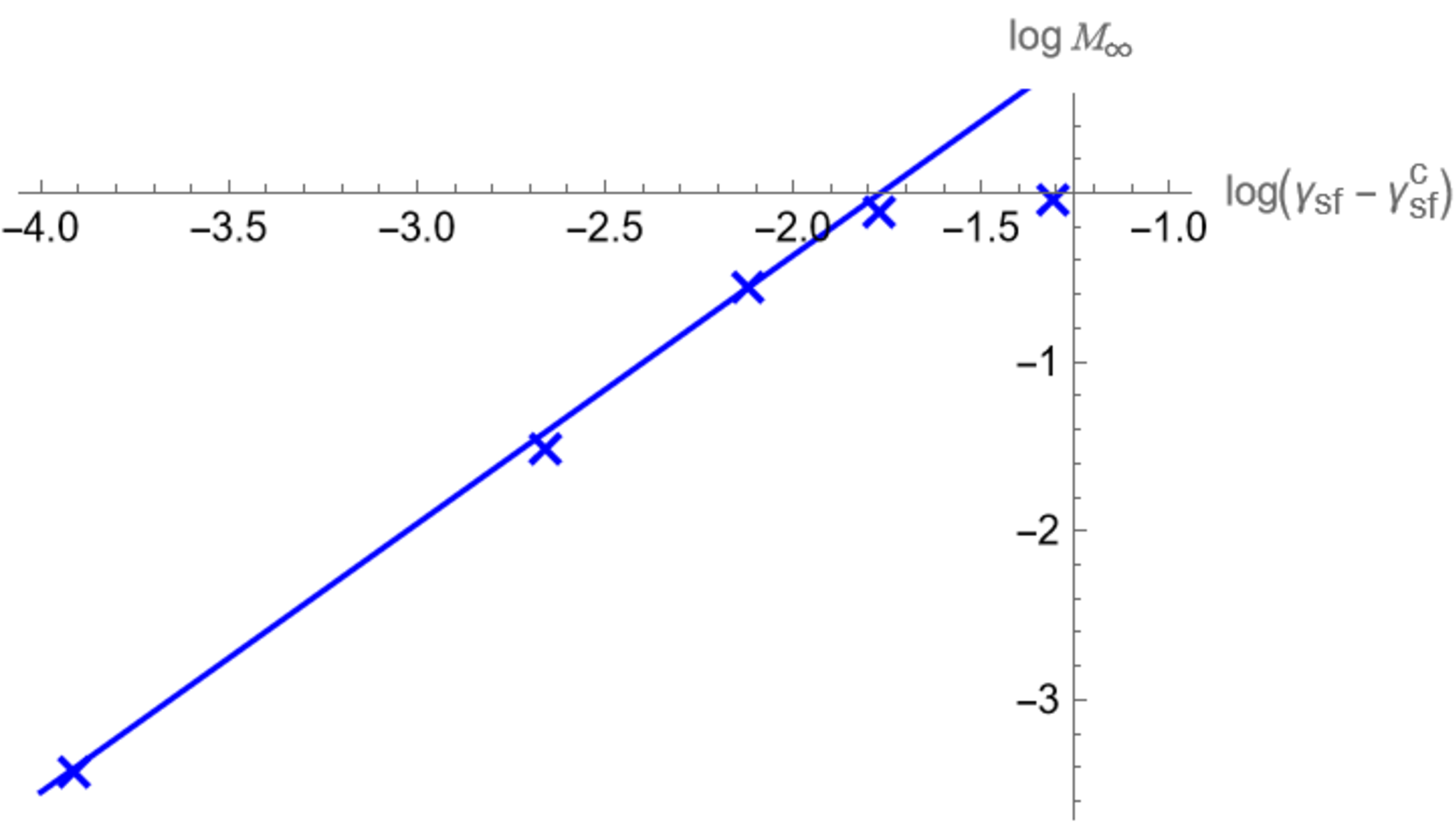}
    \subcaption{Fitting with $\gamma_{\rm sf}^{\rm c}=0.63$ and $\beta=1.59$
    for $M_{\infty}$.}
  \end{minipage}
  \\
\vskip 0.3cm
  \begin{minipage}[b]{0.50\linewidth}
    \centering
    \includegraphics[keepaspectratio, scale=0.3]{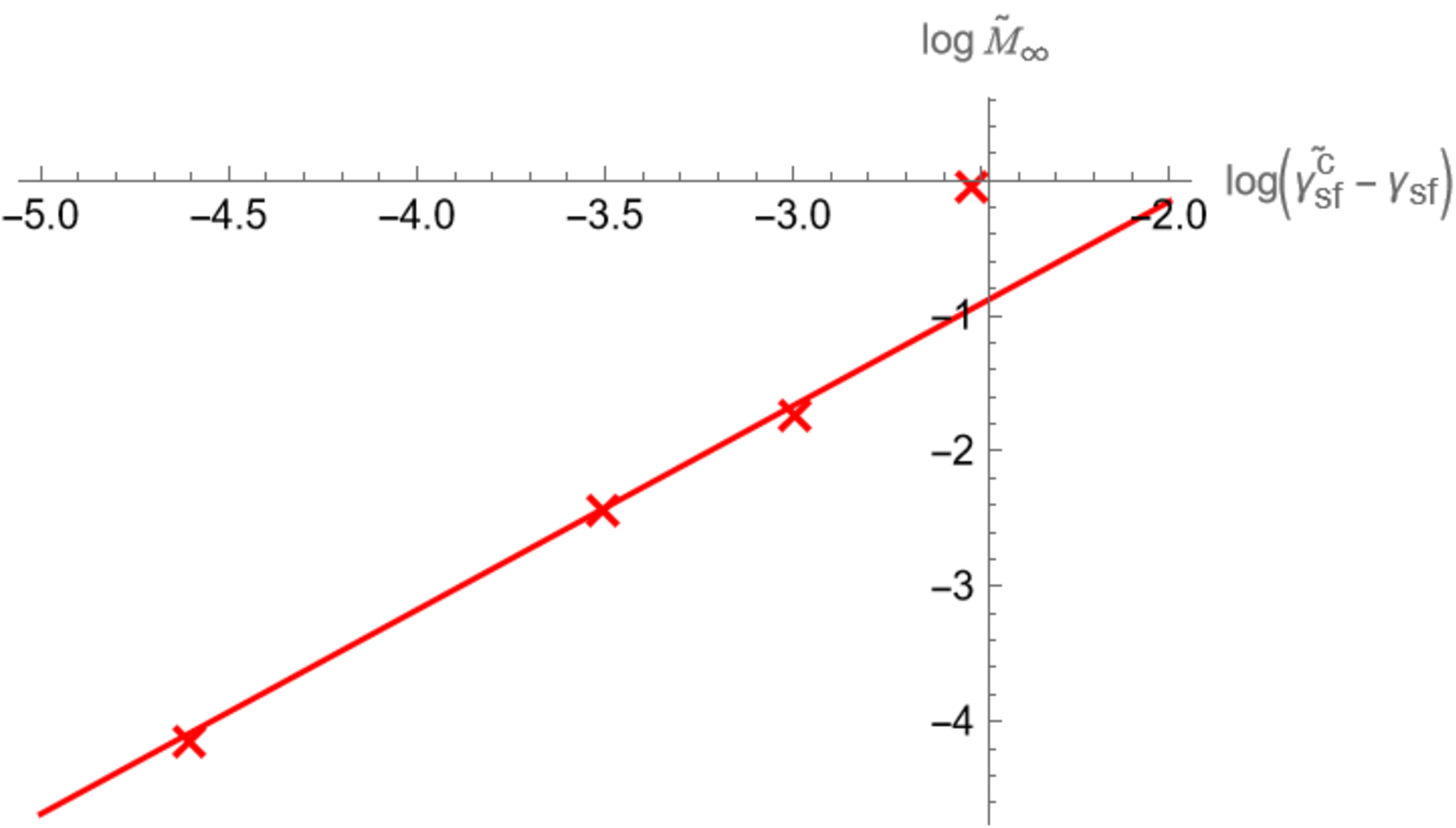}
    \subcaption{Fitting with $\widetilde{\gamma_{\rm sf}^{\rm c}}=0.08$ 
    and $\widetilde{\beta}=1.51$ for $\widetilde{M}_{\infty}$.}
  \end{minipage}
  \caption{Log-log plots of the order parameters $M_{\infty}$ and $\widetilde{M}_{\infty}$
with respect to the deviations of the parameter
$\gamma_{\rm sf}$ from its critical values,
$\gamma_{\rm sf} - \gamma_{\rm sf}^{\rm c}$
and
$\widetilde{\gamma_{\rm sf}^{\rm c}}-\gamma_{\rm sf}$,
respectively.
The slopes provide the critical exponents
$\beta$ and $\widetilde{\beta}$.}
\label{fig:loglog}
\end{center}
\end{figure}

The results suggest 
that we can define the two types of 
critical values of $\gamma_{\rm sf}$ according to 
the two types of order parameters, $M_{\infty}$ and $\widetilde{M}_{\infty}$,
as
\begin{equation}
\gamma_{\rm sf}^{\rm c}
:= \inf \{ \gamma_{\rm sf} : M_{\infty} > 0 \},
\quad
\widetilde{\gamma_{\rm sf}^{\rm c}}
:= \sup \{ \gamma_{\rm sf} : \widetilde{M}_{\infty} >0\}.
\label{eqn:critical_values}
\end{equation}
From the $1/L$-plots, we 
evaluate the critical values as follows:
\begin{equation}
\gamma_{\rm sf}^{\rm c} \simeq 0.63, \quad
\widetilde{\gamma_{\rm sf}^{\rm c}}
\simeq 0.08. 
\label{eqn:critical_values2}
\end{equation}
See Fig.~\ref{fig:OrderParameters} (b). 

The continuous changes in $M$ and $\widetilde{M}$
at these critical points suggest that the phase transitions
are continuous. 
The critical phenomena are then described by the following
\textit{power laws} as functions of
the parameter $\gamma_{\rm sf}$,
\begin{align}
M_{\infty}(\gamma_{\rm sf})
&\simeq c_2 
(\gamma_{\rm sf}-\gamma_{\rm sf}^{\rm c})^{\beta},
\quad \gamma_{\rm sf} \gtrsim \gamma_{\rm sf}^{\rm c},
\nonumber\\
\widetilde{M}_{\infty}(\gamma_{\rm sf})
&\simeq \widetilde{c_2} 
(\widetilde{\gamma_{\rm sf}^{\rm c}}
-\gamma_{\rm sf})^{\widetilde{\beta}},
\quad \gamma_{\rm sf} 
\lesssim \widetilde{\gamma_{\rm sf}^{\rm c}},
\label{eqn:critical_exponent}
\end{align}
where 
$\beta$ and $\widetilde{\beta}$ are the
\textit{critical exponents of order parameters}
$M$ and $\widetilde{M}$, respectively 
\cite{CD98,MD99,VZ12}.
As shown in Fig.\ref{fig:loglog}, 
the power laws are observed and 
the log-log plots
provide the values
\begin{equation}
\beta \simeq 1.6, \quad
\widetilde{\beta} \simeq 1.5.
\label{eqn:beta}
\end{equation}

\SSC
{Concluding Remarks and Future Problems}
\label{sec:future}

At the end of this paper, we list out the 
concluding remarks and possible future problems.
\begin{description}
\item{(i)} \quad
The interactions between the ants are given by 
the pheromone field,
which evolves over time 
and is affected by the previous behaviors of 
other ants.
In this study, we have investigated such an interacting
particle system
by numerically simulating a cellular automaton model. 
What theory is relevant for describing
such temporally inhomogeneous interactions
with memory effects?

\item{(ii)} \quad
In the \textit{finite-size scaling theory} 
\cite{CD98,MD99,VZ12},
$1/L$-fitting \eqref{eqn:1/L_1} should be
refined by introducing the exponents $\nu$ 
and $\widetilde{\nu}$ as follows:
\begin{equation}
M_{\infty}=M_L + \frac{c_1'}{L^{\nu}},
\quad
\widetilde{M}_{\infty} 
=\widetilde{M}_L+\frac{\widetilde{c_1}'}{L^{\widetilde{\nu}}},
\quad \mbox{as $L\to \infty$},
\label{eqn:1/L_2}
\end{equation}
at each value of $\gamma_{\rm sf} \in [0,1]$
with \eqref{eqn:relation1}, 
where $c_1'$ and $\widetilde{c_1}'$ are 
the fitting parameters, and
$M_{\infty}$ and $\widetilde{M}_{\infty}$ are the estimated
values of the 
order parameters for an infinite system.
In this study, we have obtained 
the numerical results for only three sizes: 
$L=40$, $60$, and $80$.
Hence, it is difficult to evaluate
the values of $\nu$ and $\widetilde{\nu}$ 
using numerical fitting.
We have assumed the mean-field value 
$\nu=\widetilde{\nu}=1$ in the fitting \eqref{eqn:1/L_1}. 
The numerical results \eqref{eqn:beta} suggest
$\beta=\widetilde{\beta}=3/2$. 
A systematic study on finite-size scaling based on
large-scale simulations will be an important future
problem.
The correlation and response functions
of the present model for the group behavior
of ants should be carefully considered.
They need to understand
the critical phenomena described by the critical exponents
$\nu$ and $\gamma$ for the divergence
of the correlation length and response functions,
respectively, 
and $\delta$ for the power-law at the critical points.

\item{(iii)} \quad
An effective theory is required to describe
the phase transitions and critical phenomena
in path selection for foraging ants.
As mentioned in item (ii) above, 
we first try to establish the mean-field theory.
Fractional values of critical exponents suggest
the emergence of \textit{fractal structures} in paths of ants
at the moment when the nearly optimal paths are first established, 
as well as at the moment when the `pheromone road' vanishes
during the path selection of the ants.

\item{(iv)} \quad
By the definition \eqref{eqn:switching}, 
parameters $\gamma_{\rm sf}$ and $\gamma_{\rm fs}$
can take any values in $[0,1]$ independently.
We have assumed in the present paper, 
however, the special relation
\eqref{eqn:relation1} between these two parameters.
The present study on the global switching
of path selections shall be generalized to
the whole parameter space
$(\gamma_{\rm sf}, \gamma_{\rm fs}) \in [0,1]^2$.
As a result, we will obtain the \textit{phase diagram}
showing three types of phases;
(a) the phase in which homing particles select
the $\Gamma$-paths (\textit{i.e.}~trailing paths of the
`pheromone road')
indicated by $\widetilde{M}>0$ and $M=0$, 
(b) the phase in which homing particles
select the nearly optimal paths 
indicated by $\widetilde{M}=0$ and $M>0$,
and (c) the intermediate phase.

\item{(v)} \quad
As shown in Table \ref{table:parameters}, we have fixed
many parameters except $L, \ell$, $N$ and
$\gamma_{\rm sf}$ with relation \eqref{eqn:relation1}.
A more general and systematic study of the
proposed model is required to investigate the dependence of
the results on other parameters.
For example, if we change the parameter $\kappa$, 
the time duration of the numerical observation is changed.
Then, the transient or relaxation phenomena 
in the initial time period and the long-term behavior
of the present processes are clarified.
It is an interesting research subject
to study the precise dependence of the results 
on the evaporation rate $q$ of the pheromone,
because this parameter controls the
spatiotemporal correlations between particles
via the pheromone field.

\item{(vi)} \quad
The present study was motivated by the experiments
reported by the group of the last author (HN) 
of this paper, as briefly explained in Section
\ref{sec:experiment} \cite{Nis15}. 
In our modeling, we adopted the setting with
$\theta=90^{\circ}$ for
the turning angle of the `pheromone road'.
(See Figs.~\ref{fig:Nishimori2015} and \ref{fig:paths} again.)
The is because no global
switching phenomena in the path selection
were observed 
in the other settings with $\theta=45^{\circ}$ and
$\theta=60^{\circ}$.
In the present model, we have introduced 
other parameters $\gamma_{\rm sf}$ and
$\gamma_{\rm fs}$ (the switching parameters
between the fast and slow modes) and showed that
the global switching phenomena in path selection
are realized as continuous phase transitions
associated with critical phenomena
at critical values $\gamma_{\rm sf}^{\rm c}$
and $\widetilde{\gamma_{\rm sf}^{\rm c}}$. 
This result suggests that even if we adopt
a setting of $\theta=45^{\circ}$ or
$\theta=60^{\circ}$, we will be able to
observe such phase transitions provided that we
properly choose the parameters
$\gamma_{\rm sf}$ and
$\gamma_{\rm fs}$.
To simulate the models by setting $\theta=45^{\circ}$ or
$\theta=60^{\circ}$ efficiently, we shall
change the lattice from the square lattice
to the triangular or the honeycomb lattices.
We expect that the critical phenomena 
associated with the phase transitions
representing the global switching phenomena in path selections are \textit{universal} 
and do not depend on any details of the lattice structure. 
If so, the present study on lattice models
will provide the universal statements 
to the path selection phenomena observed
in real foraging ants.

\end{description}

\vskip 1cm
\noindent{\bf Acknowledgements} \,
The authors would like to express their gratitude to
Mitsugu Matsushita,
Frank den Hollander, 
Helmut R. Brand,
Piotr Graczyk,
Gregory Schehr, 
and 
Hirohiko Shimada 
for their valuable discussions on this work.
This study was carried out under the 
Open-Type Professional Development Program 
of the Institute of Statistical Mathematics, Tokyo
(2023-ISMHRD-7010). 
Part of this study was 
presented by AE on December 6, 2023, 
at the Institute for Mathematical Sciences,
National University of Singapore.
AE, SM, YT, and MK are grateful to
Akira Sakai and Rongfeng Sun
for organizing the three-week fruitful program 
in December 2023.
Additionally, this work received support from
the Research Institute for Mathematical Sciences,
an International Joint Usage/Research Center located 
at Kyoto University.
MK is funded by
the Grant-in-Aid for Scientific Research
(C) (No.24K06888), 
(C) (No.19K03674), 
(B) (No.23K25774), 
(B) (No.23H01077), 
(A) (No.21H04432), 
and the Grant-in-Aid for Transformative Research Areas
(A) (No.22H05105)
of the Japan Society for the Promotion of Science (JSPS).
HN is supported by
the Grant-in-Aid for Scientific Research
(B) (No.20H01871), 
and
the Grant-in-Aid for Transformative Research Areas
(A) (No.21H05293) and 
(A) (No.21H05297)
of JSPS.



\end{document}